\documentclass[journal=jacsat,manuscript=article]{achemso}

\usepackage{graphicx} 
\usepackage{amsmath}
\usepackage{float}
\usepackage{xcolor}
\usepackage{outlines}

\newcommand{\dmin}{$D^2_{min}$}

\title{Thermally activated dynamics of annealed glasses near the yielding transition under cyclic shear}

\author{Ian R Graham}
    \affiliation{Department of Physics and Astronomy, University of Pennsylvania, Philadelphia, PA 19104, USA}
\author{Paulo E Arratia}
    \email{parratia@seas.upenn.edu}
    \affiliation{Department of Mechanical Engineering and Applied Mechanics, University of Pennsylvania Philadelphia, PA 19104, USA}
\author{Robert A Riggleman}
    \email{rrig@seas.upenn.edu}
    \affiliation{Department of Chemical and Biomolecular Engineering, University of Pennsylvania Philadelphia, PA 19104, USA}

\begin{document}

\begin{abstract}
    While experiments and simulations have provided a rich picture of the dynamic heterogeneity in glasses at constant temperature or under steady shear, the dynamics of glasses under oscillatory shear remain comparatively less explored. Recent work has shown that oscillatory shear protocols can embed a ``memory'' into a glass's structure, whereby the material will exhibit dynamics that are encoded by the oscillatory shear protocol applied. However, most of the computational work studying the memory effect has been performed in the zero temperature limit, and the effects of thermalization are poorly characterized. In this work, we use nonequilibrium molecular dynamics simulations to study the dynamics of a model two-dimensional glass former at low, non-zero temperatures under oscillatory shear. While we show that the systems' dynamics are independent of sample preparation for either small or larger strain amplitudes, the dynamics become distinct near the yield point when the deformation is applied at finite temperature. We then characterize the dynamic heterogeneity using two metrics, one derived from the vibrational modes and one that exploits machine learning to identify regions prone to rearrangement. This analysis provides evidence that the dynamics below and above yield emerge from distinct structural origins that may be important for developing improved constitutive models that can predict memory in disordered solids.
\end{abstract}

\maketitle

\section{Introduction}

Many materials found in nature are disordered. From the soil beneath us, to glass, paint, and plastics, these materials contain constituents that are arranged in structurally disordered configurations \cite{larson99, nagel2017, jerolmack-daniels2019}. Such disordered systems find many uses in technology such as coatings, screens, biomedicine, and manufacturing, often because they are strong (high yield stress) \cite{Bonn_Yielding, Falk_Langer_RevModPhys}. However, these systems suffer from low toughness; that is, the material is unable to resist large deformations without fracturing \cite{Ediger_PhysToday_2016, Falk_Langer_RevModPhys}. At the macroscopic level, the fracturing processes is characterized by irreversible (macroscopic) deformation; at the constituent (e.g., particles in colloidal system) level, fracture is accompanied by local rearrangements that significantly dissipate energy (or stress) due to changes in neighbors or bonds.  A challenge in understanding yielding and failure in glassy systems is that these behaviors are not easily intuited from simple physical laws or relations. Distinct from crystals, where periodic structure coupled with dislocation theory is vital to predicting local rearrangements \cite{rodriguez_sixty_1996}, there is not an analogous or obvious structural parameter to quantify disorder in amorphous solids, with early proposed parameters such as the local free volume falling to capture the heterogeneous dynamics \cite{debenedetti_supercooled_2001, chaudhari_edge_1979, gilman_metallic_1975, widmer-cooper_free_2006, widmer-cooper_predicting_2006}. The difficulty in characterizing and predicting the behaviors of disordered materials primarily lies in the lack of any easily identifiable structural parameter that correlates with local deformation or rearrangements \cite{cubuk_structure-property_2017, cubuk_structural_2016, richard_predicting_2020}. 

Historically, it has been understood that disordered materials possess features in their vibrational density of states (VDoS) distinct from those generically found in ideal elastic solids or crystals \cite{lerner_low-energy_2021}. For instance, while Debye theory predicts that for an ideal elastic solid (in 3D) the distribution of low-frequency VDoS scales as $\sim \omega^2$, where $\omega$ is the frequency, many disordered solids instead show a $\sim\omega^4$ scaling that terminates in a boson-peak at intermediate $\omega$ before merging back with the high-frequency phonon spectrum \cite{lerner_boson-peak_2023}. Both the low-frequency spectrum and boson-peak have long been hypothesized to be critical in explaining many of the complex mechanical, thermodynamic, and acoustic properties of disordered materials \cite{phillips_structure_1978}. The material properties that are affected by these non-phononic modes are not just due to the modification of the VDoS, but more importantly are due to the structure of these excitations; these excess modes of disordered solids are found to be highly localized \cite{schober_localized_1991}. These excitations have been suggested to be good candidates to identify with shear-transformation zones (STZs) \cite{falk_dynamics_1998}, the elementary irreversible rearrangement regions of disordered materials. Indeed, numerical simulations have shown the correlation of numerous quantities derived from these low-frequency modes with the irreversible rearrangements of supercooled liquids and athermally sheared granular packings \cite{widmer-cooper_irreversible_2008, manning_vibrational_2011, tong_order_2014}. But as larger system sizes are analyzed, issues in the application of these modes arise due to their hybridization with phonon modes \cite{gartner_nonlinear_2016}. To address this challenge, several techniques have been proposed to either remove phononic contributions from normal modes \cite{gartner_nonlinear_2016, zylberg_local_2017}, or construct non-linear modes with custom cost functions that may map better to localized STZs \cite{richard_simple_2021}. Under steady shear simulations the prior approach, which we call here the \emph{soft modes}, was found to correlate well with rearrangements when compared to purely structural measures \cite{richard_predicting_2020}. However, accurately measuring the material soft modes remains quite difficult in experiments, which limits their applicability.

In a separate vein, significant progress has been made in defining structural measures that are correlated with rearrangement dynamics under shear and thermal activation, particular using machine-learning methods \cite{cubuk_identifying_2015,cubuk_structure-property_2017, richard_predicting_2020, yang_role_2022}.  One of the applications of this methodology is through a quantity named \emph{softness}, which is constructed to identify defects in disordered systems, analogous to dislocations in crystalline systems, that are strongly correlated with rearrangements. Softness is derived from a weighted sum of coarse samples of the local radial distribution function of a particle \cite{cubuk_identifying_2015, schoenholz_structural_2016, cubuk_structure-property_2017}. The weights are obtained from the optimization of a linear support vector machine (SVM) model, and a linear combination of these weights are used to identify rearranging from non-rearranging particles. In supercooled liquids, for example, this method can identify whether a given particle is likely to \emph{hop} within the typical duration of glassy rearrangements \cite{Ma_SoftColloids_PRL}. Of particular note among the results using softness, the commonly observed non-Arrhenius dynamics of glasses becomes Arrhenius when measured in groupings of softness \cite{cubuk_identifying_2015, schoenholz_structural_2016}. This implies that softness is effectively capturing structural information that is related to the local energy barriers to rearrangement. Following this work, many softness-based analyses were performed on systems from dense colloids, to bubble rafts, oligomer pillars, and glassy thin films \cite{yang_role_2022, sussman_disconnecting_2017, cubuk_structure-property_2017}. Similar to the soft modes, softness correlates well with rearrangements under steady shear \cite{richard_predicting_2020} and has even been used to develop new, structurally based constitutive models\cite{zhang_PhysRevRes_2022}. 

Recently, there has been significant interest in the relationship between microstructural dynamics and macroscopic behavior (e.g., yielding and plasticity) in athermal glasses under oscillatory shear\cite{keim_yielding_2013, keim_multiple_2013, keim_mechanical_2014, adhikari_memory_2018, fiocco_memory_2015, fiocco_encoding_2014, leishangthem_yielding_2017, lavrentovich_period_2017, yeh_glass_2020, regev_reversibility_2015, regev_irreversibility_2017, regev_topology_2021, regev_reversibility_2015-1, regev_onset_2013, szulc_cooperative_2022, kumar_mapping_2022, szulc_overlapping_2024, reichhardt_reversible_2023, keim_memory_2019, keim_multiperiodic_2021, galloway_relationships_2022, galloway_scaling_2020, lawrencegalloway_quantification_2020, teich_crystalline_2021, fiocco_oscillatory_2013}. Below the yielding transition, these systems readily form limit-cycles in configuration space, where after a full cycle of oscillatory shear the system returns to the starting configuration despite nonaffine displacements emerging during the cycle. Stemming from this has been increased interest in understanding the origin of this behavior and characterizing the irreversible plastic, reversible elastic, and reversible plastic events that occur within the cycle \cite{galloway_scaling_2020}. There has been some success in connecting the observation of these microscopic reversible dynamics and the structural evolution of the system to the bulk oscillatory rheology in 2D colloidal experiments \cite{galloway_relationships_2022}. Only a few studies, however, have explored the effect of thermal activation on memory formation under oscillatory shear dynamics \cite{majumdar_memory_2023}. While it is known that the sample (thermal) preparation history plays an important role in the dynamics of glasses \cite{kovacs_transition_1964, cerrada_physical_2000, mckenna_aging_1991, mckenna_physical_1984, lee_mechanical_2010, bennin_rejuvenation_2020}, preparation dependence studies are often restricted to an athermal quasistatic shear (AQS) protocol \cite{yeh_glass_2020, leishangthem_yielding_2017}. Previous works that have explored the effects of thermal activation on reversible dynamics under oscillatory shear have only used one preparation temperature point \cite{majumdar_memory_2023}, and the role of different preparation histories remains relatively unexplored.

In this work, we investigate the dynamics of a 2D Kob Andersen glass under oscillatory shear and thermal activation following distinct sample preparation histories. We find that the steady-state particle rearrangement dynamics for samples of varying history are nearly the same far below yield and far beyond the yielding transition, consistent with results of reversible limit cycles and glassy rejuvenation beyond yielding in AQS simulations. However, across the yielding transition, we observe distinct and diffusive dynamics that depends on the thermal history in the presence of thermal activation; to our knowledge this history dependence has not been previously reported in the literature. We analyze our results by taking a closer look at the correlation of the individual particle dynamics to the structural descriptors described above, softness and the soft modes. We find that machine learned softness and the soft modes show distinct advantages in different regimes relative to the yielding transition. These results suggest fruitful directions to constructing microscopic quantities of the structure that correlate better with oscillatory shear dynamics in all shear regimes, and may aid in developing better constitutive models of reversible and irreversible plastic deformation under oscillatory shear.

\section{Methods}

\subsection{Simulation}

We performed simulations of a 2D Kob-Andersen mixture with a 60:40 ratio of type A to B particles with Lees-Edwards periodic boundary conditions \cite{kob_scaling_1994}. The Lennard-Jones interaction parameters are set to $\epsilon_{AA}=1.0$, $\epsilon_{AB}=1.5$, $\epsilon_{BB}=0.5$, $\sigma_{AA}=1.0$, $\sigma_{AB}=0.8$, and $\sigma_{BB}=0.88$. To obtain initial samples, we first performed gradual quench simulations from the high-temperature liquid state into the supercooled liquid. Measuring the cage-breaking relaxation time $\tau_\alpha$ against the inverse temperature $1/T$, we estimate the VFT temperature, $T_{VFT} \approx 0.2$, by fitting the relaxation time data to the VFT equation. We interpret $T_{VFT}$ as the end of the glass forming regime, and the observed $T_g$ is typically 20-30\% higher in temperature \cite{dudowicz_entropy_2006, dudowicz_glass_2005}. We simply use $T_{VFT}$ to identify the temperature range where we expect our samples to fall out of equilibrium, and the precise definition of $T_g$ does not play a role in the results presented below.

Before running shear experiments, we generate samples with two different preparation protocols. The first is a glass rapidly quenched from the liquid state at $T=1.5$, which we call here the high-temperature liquid (HTL) samples. The second is a liquid that has been gradually cooled to $T=0.3 \approx 1.5T_{VFT}$ (while maintaining equilibrium), which we call the gradual quench (GQ) samples. All samples have a 2D number density $\rho=1.2$ with a total particle count of $N=2^{15}=32,768$. Six randomly initialized replicas are generated for each preparation history, for a total of 12 samples. All simulations were performed using the HOOMD-blue simulation library \cite{anderson_hoomd-blue_2020}.

We perform both thermal and athermal oscillatory shear simulations from the prepared samples. The thermal simulations are performed within the NVT ensemble using a Nos\'e-Hoover thermostat with a coupling constant $\tau=100\Delta t$, where $\Delta t=0.005$ is the simulation time step size. We explored thermostat temperatures ranging from as high as $T=\frac{1}{2}T_{VFT}$, down to $T=\frac{1}{100}T_{VFT}$. We primarily analyzed samples where the oscillatory shear was a sinusoidal function driven at a period of $\tau=1,000\tau_{LJ;A}$. This timescale was chosen so to be considerably larger than the duration of rearrangements, which in the higher temperature supercooled simulations is no more than $\tau_{hop} \approx 10\tau_{LJ;A}$ at $T=0.3$. The athermal simulations follow an AQS protocol by modifying the box tilt of the system in incremental steps of $\Delta \gamma = 10^{-3}$. After each strain step, the configurations are minimized to the nearest inherent structure using the FIRE algorithm, while the system box is fixed \cite{bitzek_structural_2006}.

\subsection{Softness}

Softness is trained identical to earlier work, where only pair-wise, radial structure functions (RDFs) are used \cite{graham_exploring_2023}. These structure functions follow a Behler-Parrinello style \cite{behler_generalized_2007}, approximating a Gaussian-smearing of the local radial distribution function centered at a particle
\begin{equation}
    \mathcal{G}_{K}(\mu_j) = \sum_{k \in \{\text{neigh}\}}{e^{\frac{-(r_k-\mu_j)^2}{2{\Delta \mu}^2}}}
\end{equation}
where the sum over $\{\text{neigh}\}$ is of all neighbors $k$ in where $r_k \in [\mu_j-3\Delta \mu,  \mu_j+3\Delta \mu]$. A smearing width $\Delta \mu=0.1$ is used. There are 54 structure functions that we use as input in the SVM model, which contain the Gaussian smeared estimates of $g(r)$ at positions $\mu_i \in [0.3, 0.4, ..., 2.9, 3.0]$. We also separate the RDFs by the particle species $K\in\{A, B\}$. We train A and B type particles separately, generating two unique hyperplanes.

Particles are grouped by measuring thermal hops\cite{smessaert_distribution_2013, smessaert_structural_2014} using $p_{hop}$ over an interval $\Delta t = 10\tau_{LJ;A}$ from samples equilibrated to a temperature $T=0.3\approx1.5T_{VFT}$. $p_{hop}$ is defined as
\begin{equation}
    p_{hop}(i,t) = \sqrt{\langle(x_i - \langle x_i \rangle_{B} )^2 \rangle_A \langle(x_i - \langle x_i \rangle_{A} )^2 \rangle_B} ,
\end{equation}
where $A$ is the interval $[t-\Delta t,t]$ and B is the interval $[t, t+\Delta t]$. Particles with $p_{hop}>0.2$ are labeled \emph{soft}, and those with $p_{hop}<0.05$ are labeled \emph{hard}. An SVM is then trained to predict whether a particle belongs to one of these two classes, and the distance to the resultant hyperplane is the particle \emph{softness}. Similar to prior work, our softness is trained on rearrangements where the box is fixed, i.e., no shear has been applied \cite{yang_role_2022}. In the analysis that follow, we take this softness trained on the supercooled liquid data and apply it to rearrangement under oscillatory shear, and temperatures well into the glassy regime.

\subsection{Non-affine deformation: $D^2_{min}$}

To capture the non-affine displacements of particles, we compute the quantity $D^2_{min}$ \cite{falk_dynamics_1998}. For any given particle with a set of nearest neighbors $\{\text{neigh}\}$, the $D^2_{min}$ is computed as
\begin{equation}
    D^2_{min}(t, t') = \sum_{j\in\{\text{neigh}\}} |\boldsymbol x_{ij}(t') - \boldsymbol \epsilon \cdot \boldsymbol x_{ij}(t)|^2,
\end{equation}
where $\boldsymbol x_{ij}(t)$ is the relative displacement between particles $i$ and $j$ at a time $t$, and $\boldsymbol \epsilon$ is the best-fit linear transformation of the relative displacements of neighbors from time $t$ to $t'$. To compute $D^2_{min}$ between any two frames of the simulation, we take the earliest of the two and compute the neighbor list, taking the first 10 neighbors sorted by distance from the central particle as the set $\{\text{neigh}\}$. We have also verified that extending this to the 20 nearest neighbors does not impact our results. 

In our work here, we are interested in the rearrangements that occur over longer timescales, particularly those related to the time of a shear cycle. We look at two forms of the $D^2_{min}$, peak-to-peak ($D^2_{min;ptp}$) and stroboscopic ($D^2_{min;strob}$). The peak-to-peak $D^2_{min}$ captures rearrangements that occur between the largest strain magnitudes within the cycle (i.e., $\gamma=\pm\gamma_{max}$), while stroboscopic $D^2_{min}$ captures the rearrangements that remain after the full shear cycle is complete.

\subsection{Soft modes}

To compute the soft modes, we first compute the Hessian of the system, defined as
\begin{equation}
    \mathcal{H} = \frac{\partial^2U}{\partial \boldsymbol x \partial \boldsymbol x}.
\end{equation}
We use a sparse Hermitian matrix solver employing the Lanczos algorithm to compute the lowest 1024 non-zero eigenvalues and eigenmodes. In practice, we find that only the first 256–512 modes contribute strongly to the correlations shown in the results.  Using only the unfiltered modes, the vibrality of particle $i$ may be computed as a sum over the available modes as
\begin{equation}
    \Psi_i = \frac{1}{n_k}\sum^{n_k}_k \frac{|\psi^k_i \cdot \psi^k_i|}{\omega^2_k}
\end{equation}
where $n_k$ is the number of modes, $\psi^k_i$ is the subcomponent of the $k$th eigenvector relating to particle $i$, and $\omega_k$ is the $k$th eigenvalue.

To filter the normal modes, we require the third-rank anharmonic tensor, which follows the computation of the Hessian as
\begin{equation}
    U''' =  \frac{\partial^3U }{\partial \boldsymbol x \partial \boldsymbol x \partial \boldsymbol x}.
\end{equation}
The asymmetry and sparsity of $U'''$  allows for a compressed representation that only requires storage that grows with the number of interacting neighbors. With the anharmonic tensor terms computed, we can contract the tensor with the normal modes on the fly, and compute the soft modes similarly to the vibrality,
\begin{equation}
    \phi^k = U''' : \psi^k \psi^k
\end{equation}
\begin{equation}
    \mathcal{M}_i = \frac{1}{n_k}\sum^{n_k}_k \frac{|\phi^k_i \cdot \phi^k_i|}{\omega^2_k}
\end{equation}

\begin{figure}[H]
    \centering
    \includegraphics[width=1.0\linewidth]{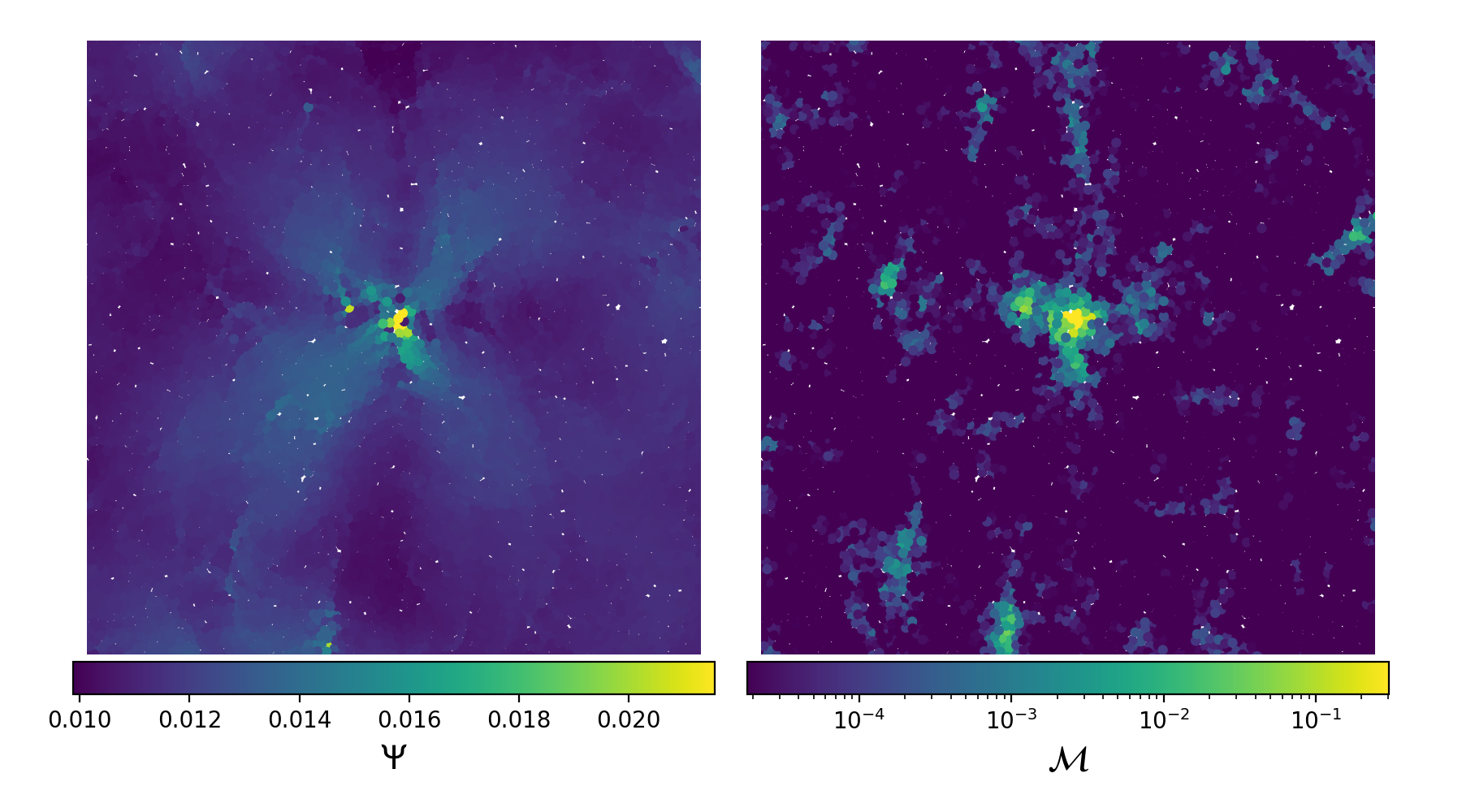}
    \caption{
    (left) Vibrality $\Psi$ computed with the first 16 non-zero modes as a demonstration. A quadrupolar far-field structure is observed along with a disordered core. (right) By contracting each mode with the anharmonic tensor, phononic and Eshelby far-fields alike are removed, leaving the disordered cores. The filtered output is the soft modes $\mathcal{M}$. Note the difference in log and linear scale bars due to the filtering.
    }
    \label{fig:3}
\end{figure}

In Figure \ref{fig:3} we show a snapshot of the vibrality $\Psi_i$ computed from the lowest 16 eigenmodes with a non-zero frequency, as well as the soft modes $\mathcal{M}_i$. The vibrality field has a long-range, quadrupolar elastic field emanating from the primary displacements in the center, while the soft modes field highlights the localized displacement field in a few clusters. Only the first 16 modes are used in this demonstration, as adding more high-frequency modes will suppress the signature of the Eshelby inclusion in the vibrality.

\section{Results}

\subsection{Diffusive dynamics below AQS yielding}

\begin{figure}[H]
    \centering
    \includegraphics[width=1.0\linewidth]{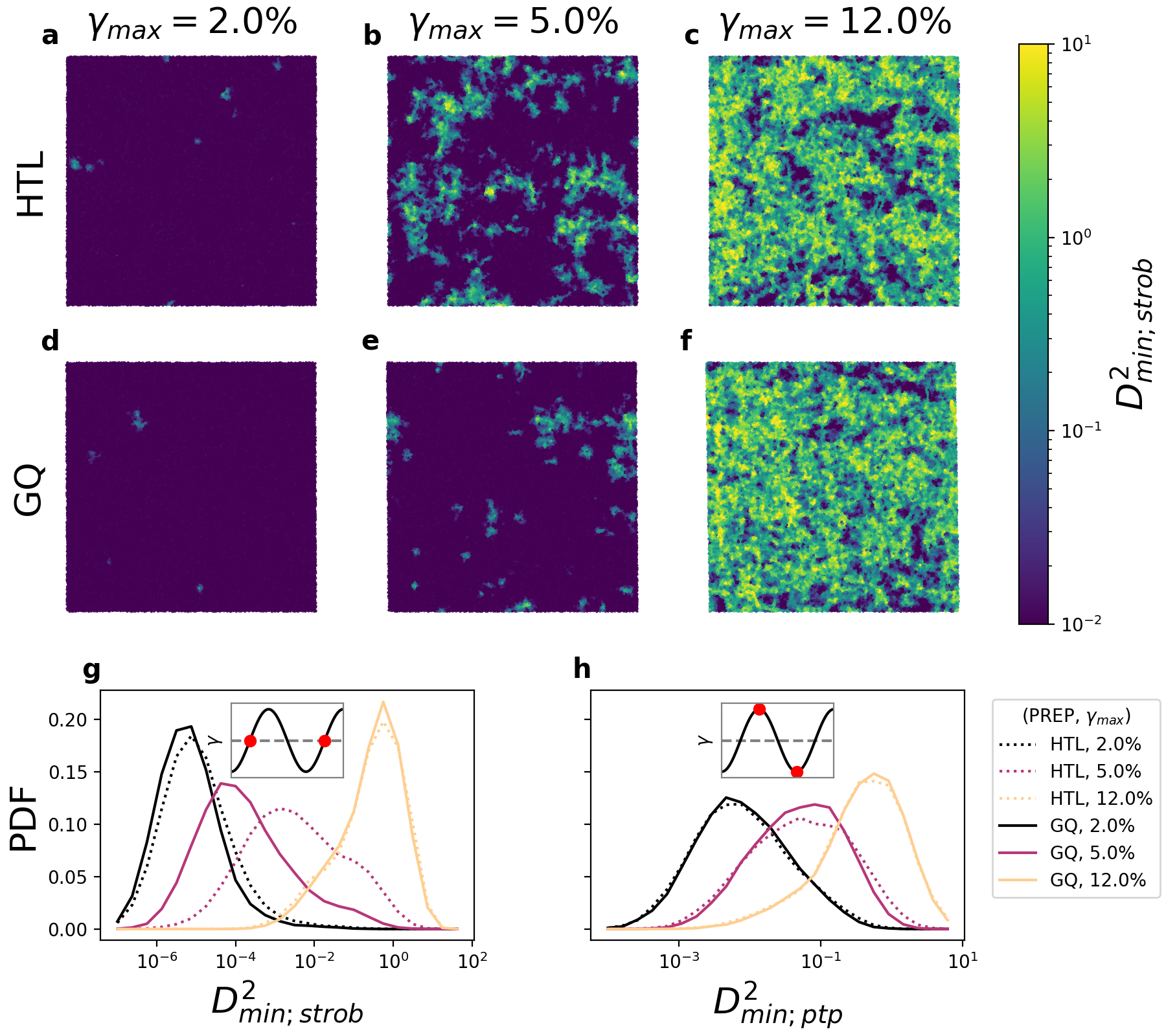}
    \caption{
    (upper) Heat maps of $D^2_{min}$ computed between adjacent stroboscopic frames of oscillatory shear. Increased $D^2_{min}$ corresponds to more significant plastic and irreversible rearrangements that have occurred in a shear cycle. The frames shown are taken 20 cycles into the simulations, where much of the transient activity has died down. The top row of figures shows results for the high-temperature liquid samples, and the middle row for the gradually quenched samples. (a, d) The left most snapshots show the simulations at $\gamma_{max}=2\%$, where rearrangements appear are localized, sparse and  independent of preparation. (b,e) $D_{min}^2$ heat maps of the simulation sheared at $\gamma_{max}=5\%$, just above the yielding transition of the HTL samples. Considerably more irreversible rearrangements are observed in the HTL samples compared to GQ. (c, f) Results from $\gamma_{max}=12\%$ well above yielding, we see similar activity in both. (g, h) PDFs of stroboscopic (strob) and peak-to-peak (ptp) $D^2_{min}$ taken from the above simulations. With $\gamma_{max}=5\%$,   $D^2_{min}$ has a much broader distribution in the HTL samples, while nearly no differences are observed in the peak-to-peak $D^2_{min}$.
    }
    \label{fig:1}
\end{figure}

First, we examine the system dynamics during oscillatory shear where the system temperature is set to $T=\frac{1}{100}T_{VFT}$. In Figures \ref{fig:1}a-f, we show heat maps of $D^2_{min}$ computed over adjacent stroboscopic frames of the oscillatory shear protocol ($D^2_{min;strob}$) after the initial transient dynamics have died down. From these images, we can see that for strains below ($\gamma_{max}=2\%$) and far above ($\gamma_{max}=12\%$) the yield point, there are few differences in the displacement fields. However, close to yield $\gamma_{max}=5\%$, there is significantly more steady-state activity in the HTL sample compared to the GQ system. This is further supported by the probability distribution functions (PDF) of $D^2_{min}$ in Figures \ref{fig:1}g \& e, where only the samples with $\gamma_{max}=5\%$ show appreciable differences in the distributions between the two preparation protocols. Interestingly, only the stroboscopic measure of $D^2_{min}$ shows a large difference, while the peak-to-peak measure is still nearly indistinguishable in the two. 
From the similarity of the distributions of $D^2_{min;stob}$ and $D^2_{min;ptp}$ between our preparations, it appears average dynamics related to these escapes does not strongly depend upon the initial thermal preparation, at least before yielding is reached, and our results are largely in accord with previous work\cite{majumdar_memory_2023} showing systems can escape limit cycles with small thermal activation. 

\begin{figure}
    \centering
    \includegraphics[width=1.0\linewidth]{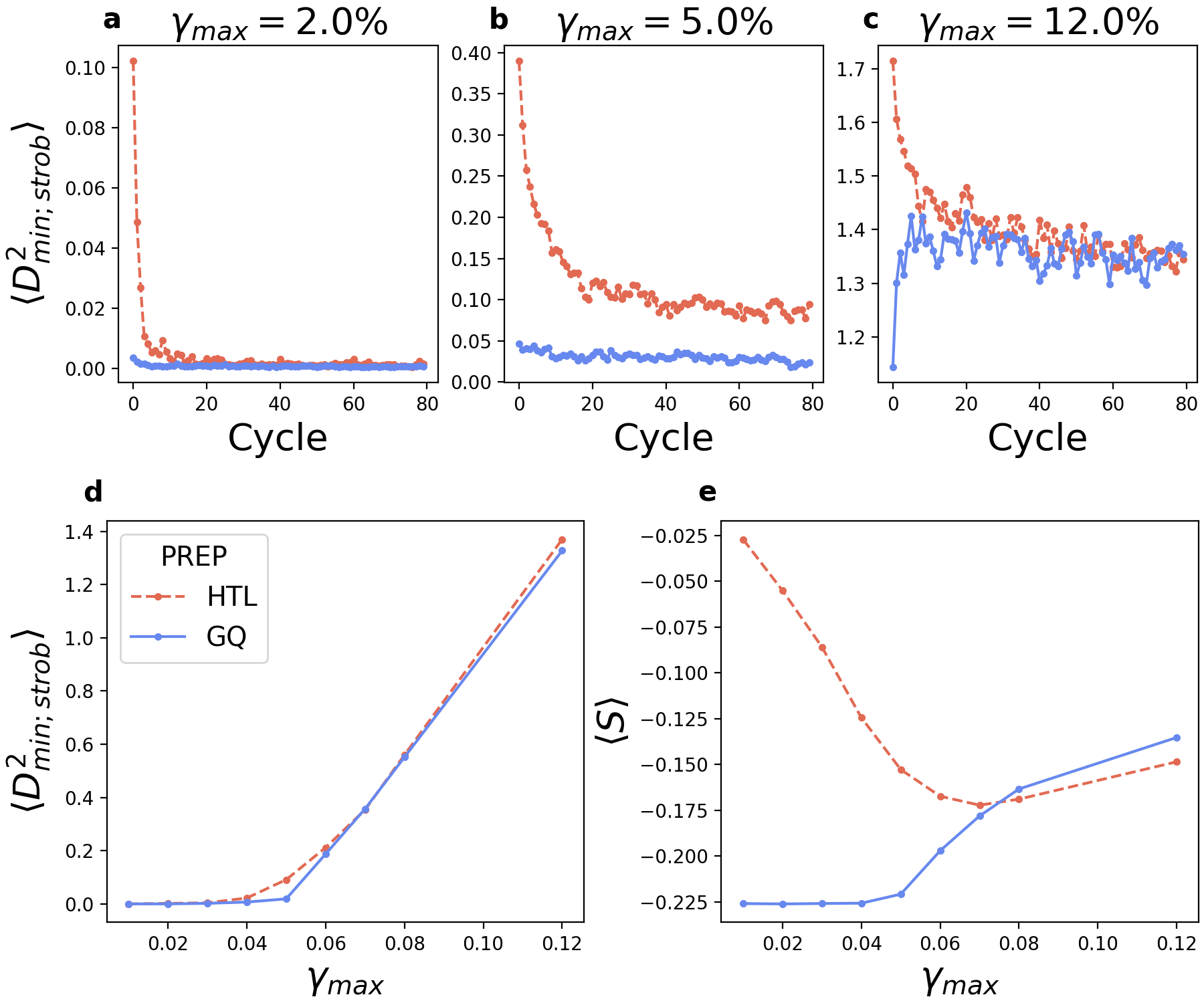}
    \caption{(a-c) Average $D^2_{min;strob}$ in a cycle plotted against the cycles since the initial preparation. Plots span from below to above yield. Steady-state dynamics develops within 80 shear cycles. GQ samples are plotted in solid blue, and HTL samples in dashed red. The GQ samples show a minor transient decay below yielding, and nearly no change at yielding. On the other hand, we see strong transient decays from at all shears in the HTL samples. (d) Average $D^2_{min;strob}$ measured after 80 cycles of shear. Microscopic yielding is clearly observed in both samples, and differs based open the preparation. (e) Average softness measured after 80 cycles of shear. Microscopic yielding transition coincides with an increase in average softness observed in GQ samples.}
    \label{fig:transient}
\end{figure}

Examples of the strain-cycle dependence of $D^2_{min}$ are shown in Figure \ref{fig:transient}a-c, which help to quantify the observations from Figure \ref{fig:1}. For the strain amplitudes below (Fig.~\ref{fig:transient}a) and above yield (Fig.~\ref{fig:transient}c), after an initial transient the values of \dmin track each other regardless of sample preparation protocol. In contrast, the thermal histories remain distinct out to 80 strain cycles for strain amplitudes near yield (Fig.~\ref{fig:transient}b). Figure \ref{fig:transient}d shows how \dmin increases as $\gamma_{max}$ increases; while the yield point remains approximately the same for the two samples, the yielding transition is much sharper in the GQ sample compared to the HTL, which has a more gradual increase in \dmin. It is expected that glasses prepared using slower cooling rates will the increase strain of the yielding transition, and a brittle response may then be observed \cite{yeh_glass_2020} with sufficient low-temperature annealing. The increase in the value of $\gamma_{max}$ where $\langle$\dmin$\rangle$ begins to increase is consistent with this effect.

In Figure \ref{fig:transient}e we can see that the average softness of the GQ and HTL samples show different behaviors below yield, with a flat profile for the GQ samples and a steadily decreasing average softness with HTL samples as $\gamma_{max}$ is increased. These results resemble simulations that explored the dependence of thermal preparation on yielding under an oscillatory AQS protocol \cite{yeh_glass_2020}, which showed a similar trend in the inherent structure energy when comparing samples that are poorly annealed and well annealed. Poorly annealed samples (similar to our HTL) show a decreasing inherent structure energy below yield, while well annealed samples (similar to GQ) maintain the inherent structure energy of their initial preparation at all strains until they yield. 

\begin{figure}[H]
    \centering
    \includegraphics[width=1.0\linewidth]{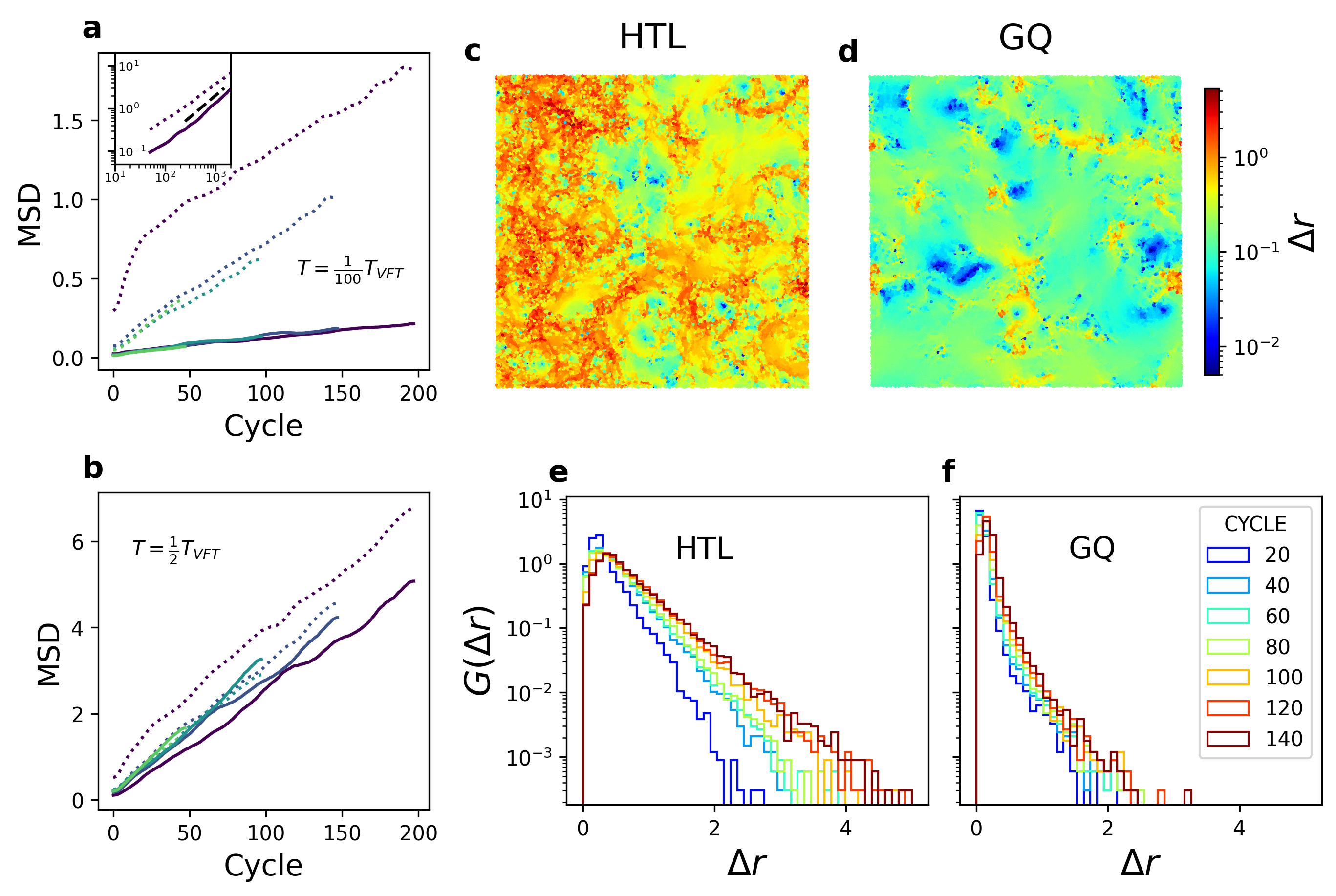}
    \caption{
    (a) Mean-squared displacement over 200 shear cycles with a temperature $T=\frac{1}{100}T_{VFT}$ and applied shear $\gamma_{max}=5\%$. Dotted lines are HTL and solid lines are GQ, colored by waiting times of 0, 50, 100, and 150 cycles. Both samples approach a steady state with approximate diffusive behavior. (a inset) MSD vs. cycle over 2000 cycles with log-log scale. The dashed line shows a diffusive trend. (b) MSD vs. cycle with a thermostat temperature $T=\frac{1}{2}T_{VFT}$ and applied shear $\gamma_{max}=5\%$. (c-d) Heat maps of particles displacements in HTL and GQ and samples over 140 cycles. In the GQ samples, particle displacements are strongly localized. (e-f) Van Hove functions over cycles for HTL and GQ samples.
    }
    \label{fig:2}
\end{figure}

Next, we analyze whether these rearrangements are associated with diffusive or caged dynamics over many cycles of shear. In Figure \ref{fig:2}a we show the MSD over stroboscopic cycles at $T=\frac{1}{100}T_{VFT}$ with an applied strain $\gamma_{max}=5\%$. Each MSD curve is with reference to a single starting cycle number 0, 50, 100, and 150 cycles into the simulation, and averaged over all simulation replicas and smoothed with a windowed-average of 10 cycles. We see that over many cycles there are significantly larger displacements in the HTL samples than the GQ ones, and this persists to over 2000 cycles of shear as shown in the inset. In Figures \ref{fig:2}c \& d we show displacement fields for the two sample histories, where one can clearly see that the HTL sample is more mobile and more homogeneous compared to the GQ sample, which only exhibits small, localized regions of high displacement. At higher temperatures that are still far below $T_g$ (shown in Figure \ref{fig:2}b), we observe that HTL and GQ samples reach similar displacements on shorter time scales. The van Hove functions in Figure \ref{fig:2}e \& f display expected behavior in the HTL samples, where the distributions have a Gaussian-like core with an exponential tail extending to larger distances, as is commonly observed in supercooled liquids and glasses \cite{kob_dynamical_1997, chaudhuri_universal_2007}. In contrast, the GQ samples have a more pronounced Gaussian core at small displacements and the exponential tail is only observable over a limited range.  The distinct dynamical behaviors that depend on sample preparation is a novel result that we do not believe has been observed in previous work, which is most commonly performed using the AQS ($T=0$) protocol. For the AQS simulations shown in Figure S1, we find that the GQ samples at $\gamma_{max}=5.0\%$ fall into a limit-cycle, and the MSD settles onto a caging plateau with increasing cycle number.

\subsection{Predicting oscillatory shear dynamics from structure of initial snapshots}

\begin{figure}[H]
    \centering
    \includegraphics[width=1.0\linewidth]{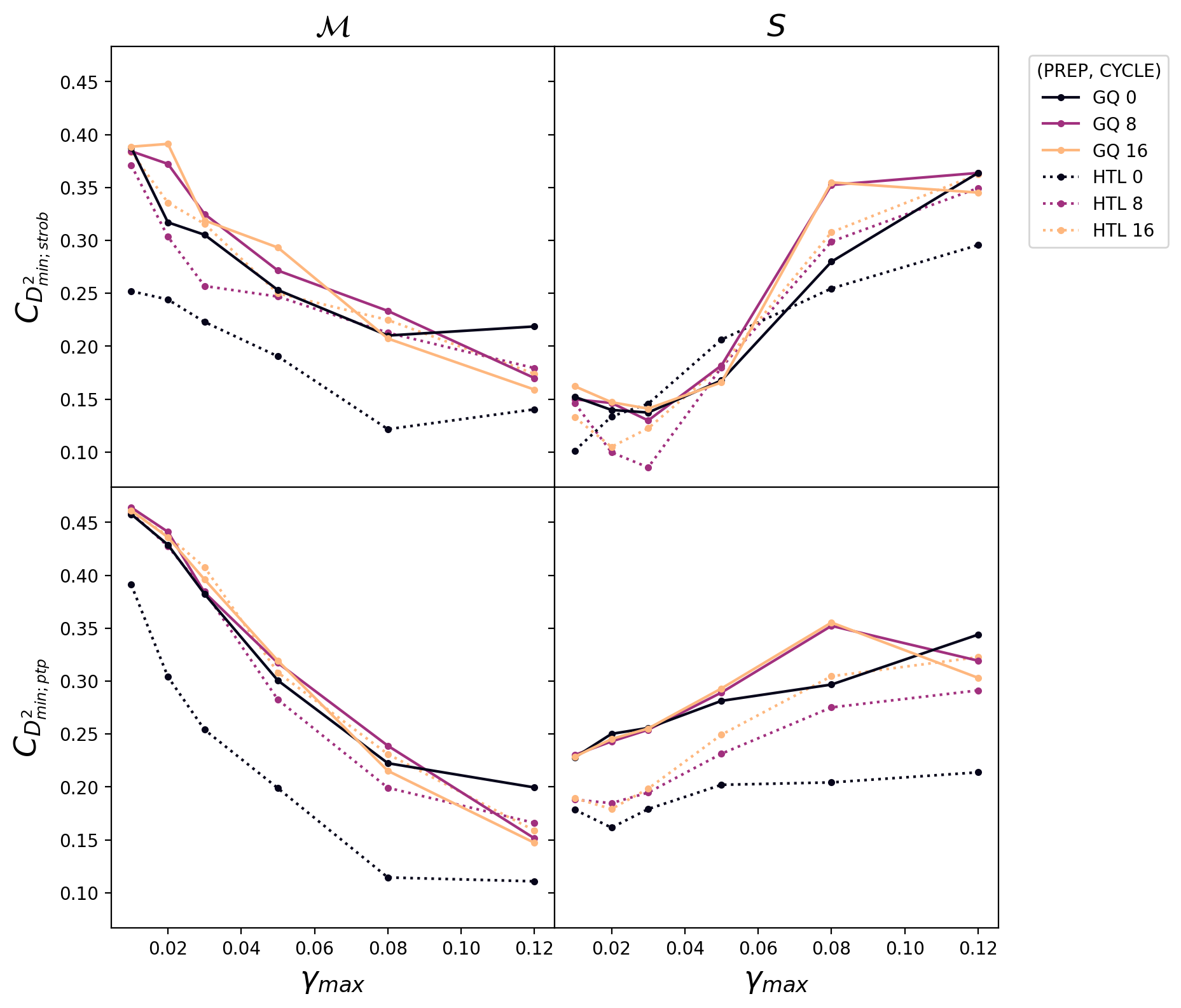}
    \caption{
    Spearman correlation coefficient for stroboscopic and peak-to-peak $D^2_{min}$. The left column shows the correlation with soft modes $\mathcal{M}$, and the right column shows the correlation with softness. Color indicates the cycle of shear, and solid lines are the gradually quenched (GQ) samples while the dotted lines are the high-temperature liquid (HTL) samples. 
    }
    \label{fig:4}
\end{figure}

In the prior section, we explored how small thermal activation can lead to qualitatively distinct dynamics that depend on sample history just beyond the HTL yielding transition. Now we turn our attention to the correlation of oscillatory shear dynamics with structure. 
In Figure \ref{fig:4} we show the Spearman correlation of the soft modes and softness with both stroboscopic and peak-to-peak $D^2_{min}$. We find that the soft modes show a stronger correlation with both stroboscopic and peak-to-peak $D^2_{min}$ below yield while softness has better correlation above yield. With softness, we see that for $D^2_{min;ptp}$, the correlation steadily increases as the applied strain is increased. In contrast, for $D^2_{min;strob}$ a weak correlation is observed in the sub-yield regime, which then improves above yield. We can conclude that softness as trained here is not as predictive of rearrangements below yield compared to the soft modes. However, once yield is surpassed and the system  dynamics and rheology are more fluid-like, softness becomes more predictive compared to the soft modes. 

The trends found with the soft modes contrast what is seen with softness. Both $D^2_{min;strob}$ and $D^2_{min;ptp}$ show moderate correlations below yield that quickly decrease with increasing $\gamma_{max}$. The moderate correlation observed with the stroboscopic $D^2_{min}$ is the most interesting, as it gives some hints to what structural quantities may be important in untangling reversible and irreversible plastic rearrangements that occur in this regime. 
In the steady state, the correlation of the soft modes with both stroboscopic and inter-cycle $D^2_{min}$ is not strongly dependent on the initial preparation, while softness shows lower correlation in $D^2_{min;ptp}$ with the HTL samples. From these results, we can conclude that the structural signatures related to rearrangements in supercooled liquids are insufficient to differentiate reversible and irreversible plastic rearrangements in the sub-yield regime. Instead, the anharmonic information of the soft modes appears to be crucial to understanding the sub-yield dynamics in disordered materials under oscillatory shear.

\section{Discussion}


Here we have studied the dynamics of a 2D model glass-forming system under oscillatory shear with distinct preparation histories that are deformed at small but non-zero temperature. Consistent with previous simulations in the $T=0$ limit using athermal quasistatic shear (AQS), we find that the steady-state dynamics of the particles in the glass are largely independent of preparation history when the maximum strain is either far below or far above the yield strain. However, for strain amplitudes near the yield point, the dynamics depend strongly on the preparation history. The samples rapidly quenched from a high temperature maintain more active dynamics compared to the gradually quenched samples, and we believe this is the first observation of this distinction.

One interpretation of the results is that when the system is sheared at sufficiently high strains, many nearby states emerge with vanishingly small energy barriers, and the system begins to diffuse among them if thermal activation is present. Resuming the shear protocol, one of the configurations is selected as the strain is reduced ($|\gamma| < \gamma^Y_{HTL}$, and as the system reaches the maximum negative strain a new group of nearby states emerge with vanishingly small barrier. From here the same process repeats. This qualitative change in the dynamics is similar to prior results of marginality in the 3D hard-sphere glass under shear \cite{jin_computer_2022}, though building a more robust connection from the present work to marginal glasses would require more detailed exploration.




Our final result showing the correlation of stroboscopic and peak-to-peak $D^2_{min}$ with two structural descriptors, softness and the soft modes. Our analysis suggests that these approaches may complement one another by becoming more predictive in different regimes relative to the yielding transition. 
We show that softness is more predictive of the rearrangements that dominate behavior beyond the yield strain, where other previous work has shown that samples' dynamics are dominated by irreversible dynamics. In contrast, the soft modes are better suited to predicting reversible deformations that occur sub-yield, where reversible nonaffine dynamics can emerge. The results suggest that anharmonicity of modes may play a critical role in determining the reversibility of rearrangements under oscillatory shear. It would be interesting to explore further non-linear formulations of quasilocalized excitations \cite{gartner_nonlinear_2016, richard_simple_2021}, apart from the anharmonic filtered normal modes used here, as the non-linear modes may map better to the shear-transformation zones (STZs) that are activated under shear. 

\section{Acknowledgements}

This work was primarily supported by the NSF through grant MRSEC/DMR-2309043.

\section{Supporting Information}
\begin{figure}[H]
    \centering
    \includegraphics[width=1.0\linewidth]{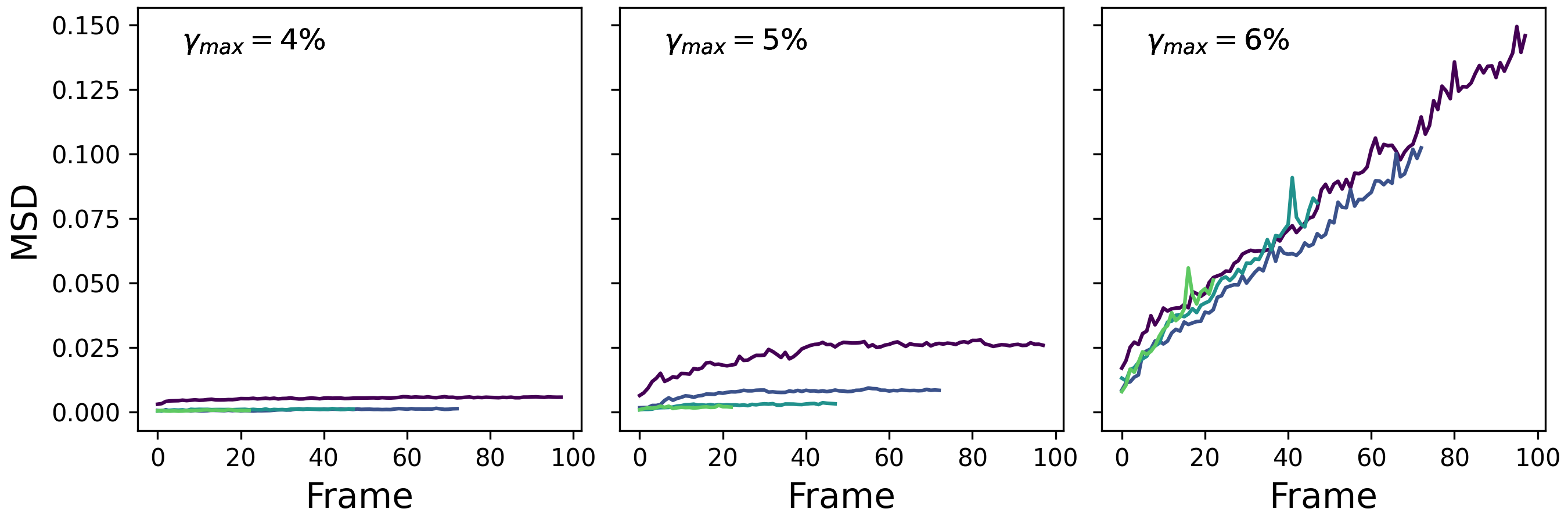}
    \caption{
    MSD of GQ samples during AQS simulations at different $\gamma_{max}$. Simulations of $\gamma_{max}=5\%$ show some mobility within the first 50 cycles, but eventually reach a plateau. Simulations of $\gamma_{max}=6$ show diffusive behavior with no apparent signs of aging over the first 100 cycles.
    }
    \label{fig:s1}
\end{figure}

\bibliography{main} 




\end{document}